\documentclass[final,5p,times,onecolumn]{elsarticle}

\usepackage{amssymb}
\usepackage{lipsum}

\journal{Physics Letters B}

\usepackage{graphicx,epsfig}
\usepackage{changes}
\usepackage{psfrag}
\usepackage{subcaption} 
\usepackage{color}
\usepackage{amsmath, amsthm, amssymb, amsfonts, amsbsy, mathrsfs}
\usepackage{url}
\usepackage{hyperref}
\usepackage{float}

\begin{document}

\begin{frontmatter}

\author[first,second]{Antonio Troisi\corref{cor}}\ead{antro@unisannio.it}
\affiliation[first]{organization={Dipartimento di Scienze e Tecnologie, Università del Sannio},addressline={Via dei Mulini 73 (Cubo)}, city={Benevento},postcode={I-82100},country={Italy}}
\affiliation[second]{organization={INFN, Sezione di Napoli Gruppo Collegato di Salerno},addressline={Complesso Universitario di Monte S. Angelo},postcode={I-80126},city={Napoli},country={Italy}}
\cortext[cor]{Corresponding author}

\author[third,fourth,fifth]{Sante Carloni}\ead{sante.carloni@matfyz.cuni.cz}\ead{sante.carloni@unige.it}
\affiliation[third]{organization={Institute of Theoretical Physics, Faculty of Mathematics and Physics, Charles University}, Prague,addressline={V Holevšovivckàch 2},postcode={CZ-180 00},city={Prague},country={Czech Republic}}
\affiliation[fourth]{organization={DIME Sez. Metodi e Modelli Matematici, Università di Genova},addressline={Via  All'Opera Pia 15}, postcode={I-16145},city={Genova},country={Italy}}
\affiliation[fifth]{organization={INFN Sezione di Genova},addressline={Via Dodecaneso 33},postcode={I-16146},city={Genova},country={Italy}}

\title{Bounce Cosmologies in Generalized Coupling Theories}

\begin{abstract}
We describe an exact solution representing a bouncing cosmology in the Minimal Exponential Measure (MEMe) model. Such a solution, obtained by means of the linearization around small values of the characteristic energy scale $q$ of the theory, has the peculiarity of representing a complete bounce model that can be used to explore quantitative processes in non-singular cosmologies.  
\end{abstract}

\end{frontmatter}

\section{Introduction}

Cosmology has witnessed remarkable advancements over the past century and the new one. The study of the large-scale structure and dynamics of the universe suggested General Relativity (GR) as the theoretical framework by which to study cosmic phenomenology from the cosmic microwave background radiation to the accelerated expansion pointed out by Ia Supernovae. However, despite these advancements, several fundamental questions remain unresolved; in particular, the nature of dark energy and dark matter still remains elusive, and we have no clear hints about the initial conditions of the universe and the ultimate fate of cosmic expansion \cite{Bamba2022,Abdalla:2022yfr}.

One often considered solution for these questions involves modifications of General Relativity (see \cite{Frusciante:2019xia,Joyce:2016vqv,Bloomfield:2012ff,Planck:2015bue} and references therein). Unfortunately, modifications of GR have the drawback of affecting even the regimes in which Einstein's theory is most successful. An interesting approach is to notice that departures between theory and observation are always present when considering non-vacuum spacetimes \cite{Berti:2015itd}, like cosmological ones. In this perspective, one can try to construct a model in which matter and spacetime (in the form of the Einstein tensor) are not linearly coupled like in Einstein's theory.  Such a theory would preserve the vacuum gravitational phenomenology but present differences when matter is present \cite{Carloni:2016glo}, potentially offering a theoretical framework for dark phenomenology. 

A concrete realization of these ideas was proposed  \cite{FengCarloni2019}, where it was recognized that such a modified theory could be considered as a bimetric theory in which the additional metric tensor is connected to the geometric one by a nondynamical coupling tensor. In the same work, the so-called Minimal Exponential Measure (MEMe) model was studied for the first time, showing that it is able to describe the onset of dark energy at cosmological scales. Unfortunately, this theory seems not to be successful in solving the problem of the flattening of galaxies rotation curve \cite{FengCarloni2021}, but it still remains an interesting theoretical framework for gravitation at cosmological scales.

Although the MEMe model has been connected to a semiclassical formulation of GR  in \cite{FengCarloni2019}, it should only be viewed as an effective representation of a more complex, and fundamental, theory of gravity. Therefore, it is important to stress that such a scheme is by no means proven to be a UV-complete theory of gravitation. Yet, it naturally offers an intriguing fresh take on different significant classical scenarios both in cosmology and astrophysics.

Perhaps one of the most riveting aspects of the MEMe model is that there exists a peculiar density at which matter effectively decouples from spacetime, and the gravitational field equations become dominated by a cosmological constant term. Given this aspect of the theory, it is normal to expect nontrivial behavior at early phases of cosmic history, and more specifically, the tantalizing possibility that this model might contain a natural bounce solution.

Hereon, we investigate this scenario. By recognizing that the MEMe model can be linearized around a small value of one of its parameters, we will be able to find an exact solution that models such a bounce behavior. We will then explore and interpret the physical properties of this solution.

Bouncing Universes have been for a long time studied as alternatives to the inflation paradigm \cite{Battefeld:2014uga,Nojiri:2017ncd,Brandenberger:2016vhg}. As inflation, they require a violation of the null-energy condition, which is not naturally occurring in GR. In most models, a bounce is achieved by introducing an additional scalar degree of freedom.  As we will see, in our case, the mechanism leading to the bounce is slightly different.  

The paper is structured in the following way. In section II, we will give the details Minimal Exponential Measure (MEMe) model, defining in particular two different field frames: the Jordan Frame and the Einstein frame, analogous to the ones of conformal and disformal transformation. In Section III, we will give the bounce solution in the Einstein frame and explore its characteristics. In Section III, we will map to the Jordan frame, and we will make a comparison of the bounce solution in the two frames. Lastly, in Section IV, we will conclude with a general discussion of the main aspects of the presented results.

We adopt here units for which $c=1$.

\section{Basic Equations and Observational constraints}
The MEMe model is a Type I minimally modified gravity theory \cite{DeFelice:2020eju,Aoki:2018brq} which is compactly defined by the action \cite{FengCarloni2019}:
\begin{equation}\label{GCA-MEMeAction}
\begin{split}
S[\phi,g,A]
=&\int d^4x \biggl\{ \frac{1}{2 \kappa}\left[R - 2 \, (\Lambda-\lambda) \right]\sqrt{-{g}} \\
&+ \left(L_{m}[\phi,\mathfrak{g}] - \frac{\lambda}{\kappa} \right) \sqrt{-\mathfrak{g}} \biggr\}.
\end{split}
\end{equation}
where $\kappa=8 \pi G$. This action is defined in terms of two metric tensors: ${g}_{\mu \nu}$ which represents the actual spacetime geometry, and $\mathfrak{g}_{\mu \nu}, $ which is the only one that couples with matter. One can imagine $\mathfrak{g}_{\mu \nu}$ to represent the geometry that one constructs on the basis of the motion of matter source, whereas ${g}_{\mu \nu}$ relies on pure geometric phenomena, like gravitational waves.  Notice that in this theory, the value of the cosmological constant is dynamic and results from the combination of two contributions: a ``naked'' term $\Lambda$ given by the observed cosmological constant and a dynamical term containing the constant $\lambda$. 

The metric $\mathfrak{g}_{\mu \nu}$ is connected to the spacetime metric $g_{\mu \nu}$ by means of a non-dynamical coupling tensor $A=A{_\sigma}{^\sigma}$
 in the following way:
\begin{equation}\label{GCA-JordanMetric}
\mathfrak{g}_{\mu \nu} = e^{(4-A)/2} \,  A{_\mu}{^\alpha} \, A{_\nu}{^\beta} \, g_{\alpha \beta} ,
\end{equation}
The field equations of the theory can be written in terms of either $g_{\mu \nu}$ or $\mathfrak{g}_{\mu \nu}$, but the first form is considerably more straightforward than the second, and we will work mainly with it. 

The equation of ``motion'' for $A{_\mu}{^\alpha}$ are:
\begin{equation}\label{GCA-ExpFEs}
\begin{aligned}
{A}{_\beta}{^\alpha} - \delta{_\beta}{^\alpha} = q \left[ (1/4) \mathfrak{T} \, {A}{_\beta}{^\alpha} - \mathfrak{T}_{\beta \nu} \, \mathfrak{g}^{\alpha \nu} \right],
\end{aligned}
\end{equation}
where 
\begin{equation}
\mathfrak{T}_{\mu \nu}=\frac{\delta L_{m}[\phi,\mathfrak{g}]}{\delta\mathfrak{g}^{\mu \nu}}
\end{equation}
is the energy-momentum tensor and $\mathfrak{T}=\mathfrak{g}^{\mu \nu}\mathfrak{T}_{\mu \nu}$, and we have defined
\begin{equation}\label{GCA-qparameter}
q=\frac{\kappa}{\lambda}.
\end{equation}
This is a key parameter of the MEMe model and regulates the deviation from GR. In particular, when $q \rightarrow \infty$ ($\lambda \rightarrow 0$), the theory reduces to GR plus a cosmological constant.

We will assume that the energy-momentum tensor for the fluid takes the form
\begin{equation} \label{GCA-EnergyMomentumPerfectFluid}
\mathfrak{T}_{\mu \nu} = \left({\rho} + p\right)u_\mu u_\nu + p \> \mathfrak{g}_{\mu \nu},
\end{equation}
where $u_\mu $ is the velocity field of matter.  Observers that are comoving with matter, and therefore with four velocity $u_\mu $, will be in free fall with respect to the metric $\mathfrak{g}^{\mu \nu}$, i.e. 
\begin{equation}
\mathfrak{g}^{\mu \nu} u_\mu u_\nu=-1
\end{equation}
However, these observers will not be free falling with respect to ${g}^{\mu \nu}$:
\begin{equation}
 g^{\mu \nu} u_\mu u_\nu = \varepsilon \neq 1\,.
\end{equation}
These results imply that observers that are free falling with respect to ${g}^{\mu \nu}$ will have four-velocity characterized by 
\begin{equation}\label{GCA-UnitFlowField}
U^\mu = {u^\mu}/{\sqrt{-\varepsilon}}.
\end{equation}
Thus we have two ``preferred'' classes of observers, one for each metric, accelerated with respect to each other. Indeed, we can define, in analogy to the case of conformal (and disformal) transformations, an ``Einstein frame'' and ``Jordan frame'' to represent the natural choice of variables for the field equations made by these two observers. More specifically, in the Einstein frame, we deal with the metric $g_{\mu \nu}$, while in the Jordan frame, we deal with $\mathfrak{g}_{\mu \nu}$. However, indices are raised and lowered in both frames using the metric $g$. 

Using the choice \eqref{GCA-EnergyMomentumPerfectFluid}, the field equation \eqref{GCA-ExpFEs} can be solved in the Einstein frame to give \cite{FengCarloni2019} 
\begin{equation}\label{GCA-AnsatzRS}
A{_\mu}{^\alpha} = {Y} \, \delta{_\mu}{^\alpha} - \varepsilon \, Z \, U{_\mu} \, U{^\alpha},
\end{equation}
where
\begin{equation}\label{GCA-ASoln}
\begin{aligned}
Y &= \frac{4 (1 - p \, q)}{4 - q \,  (3 \, p - \rho)} \\
Z &= - \frac{q \, (p + \rho) [4 - q \, (3 \, p - \rho)]}{4 \, (q \, \rho + 1)^2}\\
\varepsilon &= - \frac{16 \, (q \, \rho + 1)^2}{[4 - q \, (3 \, p - \rho)]^2}.
\end{aligned}
\end{equation}
and one obtains  that $A=A{_\sigma}{^\sigma}=4$. 

In the Einstein frame, the gravitational equations can be written as  \cite{FengCarloni2019} 
\begin{equation}\label{GCA-GEN-GFE-MEMe}
 G_{\mu \nu} +\left[  \Lambda- \, \lambda\left(1 - |A| \right)\right] \, g_{\mu \nu} = \kappa \, |A| \, \bar{A}{^\alpha}{_\mu} \, \bar{A}{^\beta}{_\nu} \, \mathfrak{T}_{\alpha \beta},
\end{equation}
where $G_{\mu \nu}$ is the Einstein tensor for $g$, $\bar{A}{^\alpha}{_\mu}$ is the inverse of $A{_\mu}{^\alpha}$ and $|A|=\det(A)$. The above field equations can be 
recast in the standard GR form as \cite{FengCarloni2021}
\begin{equation}\label{GCA-GEN-GFE-EF}
G_{\mu \nu}=\kappa \, T_{\mu \nu},
\end{equation}
where $T_{\mu \nu}$ is an effective energy-momentum tensor defined by
\begin{equation}\label{GCA-ExpTmnEffDecomp}
T_{\mu \nu} = \left(\tau_1 + \tau_2\right) \, U_\mu \, U_\nu + \tau_2 \, g_{\mu \nu} ,
\end{equation}
and
\begin{equation}\label{GCA-ExpTmnEffDecompTs}
\begin{aligned}
\tau_1 & = |A| \, (p + \rho) - \tau_2 , \\
\tau_2 & = \frac{|A| \, (p \, q - 1) + 1}{q}-\frac{\Lambda}{\kappa},
\end{aligned}
\end{equation}
with the following expression for the determinant:
\begin{equation}\label{GCA-ExpAdet}
|A| = \det(A)=\frac{256 \, (1 - p \, q)^3 (q \, \rho + 1)}{[4 - q \, (3 p - \rho) ]^4}.
\end{equation}

As one can see from the above expression, when the modulus of the energy density (or the pressure) takes values that approach $|q|^{-1}$, the coupling tensor $A{_\mu}{^\alpha}$ becomes degenerate. Consequently, the geometry corresponding to $\mathfrak{g}$ breaks down while the gravitational metric $g$ is still regular. This feature allows the construction of models in which the breaking of $\mathfrak{g}$ can be explored, or, on the other hand, the study of models that evidence critical behaviors along some peculiar specific solutions where Jordan observers and Einstein ones experience a different physics. Here, we are interested in studying bounce scenarios in the MEMe model. We will explore these solutions in both frames, giving a complete, albeit elementary, model of bouncing cosmology \cite{Nojiri:2017ncd,Brandenberger:2016vhg}.

A key element for our analysis comes from the value of the constant $q$. At present the value of $q$ has been constrained by two sources of data. The first comes from the constraint on gravitational wave speed from the kilonova event  GW170817 \cite{LIGOScientific:2017vwq,FengCarloni2019} and the second from consideration on the PPN limit of the theory \cite{Feng:2022rga}. These analyses lead to $q<0$ and the lower bound
\begin{equation}\label{GCA-qConstraint}
|q| \lessapprox 10^{-24} ~  \text{m}^3/\text{J}.
\end{equation}
As $q$ has a small value compared to almost all the other quantities in the cosmological equations, we can approximate these equations to the first order in $q$. In the following, we will use this approximation to obtain some exact and numerical solutions.

\section{Bounce Solution in the Einstein Frame}
In order to obtain cosmological solutions of Eqs.\eqref{GCA-GEN-GFE-MEMe}, let us consider a Friedmann-Lema\^{\i}tre-Robertson-Walker geometry in the Einstein frame with a line element of the form
\begin{equation}\label{FLRW}
ds^2=g_{\alpha\beta}dx^\alpha dx^\beta=-dt^2+S^2\left(t\right)\left[\frac{dr^2}{1-kr^2}+r^2d\Omega^2\right],
\end{equation}
where $k=-1,0,1$ is the spatial curvature, $d\Omega^2$ the infinitesimal solid angle and $S$ is the scale factor.
In the Einstein frame, the cosmological equations can be written as \cite{FengCarloni2019}
\begin{align}
H^2 +\frac{k}{S^2}=& \frac{256 \kappa (1-pq)^3 (q \rho +1)^2}{3q[4+q (\rho-3p)]^4}+\frac{\Lambda}{3} - \frac{\kappa}{3q},\label{FriedEq}\\
\dot{H}+H^2=& -\frac{256\kappa (p q-1)^3 (q \rho +1) [2-q(\rho+3p)]}{ 6q [4+q (\rho-3p)]^4} \nonumber\\
&+\frac{\Lambda}{3} - \frac{\kappa}{3q},\label{RayEq}
\end{align}
 where $H=\dot{S}/S$ and we have assumed a barotropic equation of state $p=w\rho$ for the fluid. On the other hand, the conservation law reads \cite{FengCarloni2019}
\begin{equation}\label{ConsLaw}
\dot{\rho}=-\frac{3 H \rho  (w+1) \left[q^2 \rho ^2 w (3 w-1)+\rho  (q-7 q w)+4\right]}{q^2
   \rho ^2 w (3 w-1)-q \rho  \left(3 w^2+13 w+2\right)+4}.
\end{equation}
The structure of previous equations suggests that in the MEMe model, the gravitation of a perfect fluid is very different from GR. Nevertheless, in this case, the three equations above are not independent, and one can be eliminated.

As mentioned before, since the parameter $q$ has a very small modulus, one can  Taylor expand all equations with respect to this quantity.  At first order in $q$, the conservation equation \eqref{ConsLaw} can be written as
\begin{equation}
   \dot{\rho }+ 3\frac{\dot{S}}{S}(w+1)\rho\left[1+\frac{3}{4 } q \rho  (w+1)^2\right]=0,
\end{equation}
which can be solved exactly to give  
 \begin{equation}\label{Sol_rho_Gen}
   \rho=\frac{4 \rho _0}{S^{3 w+3} \left[3 q \rho _0
   (w+1)^2+4\right]-3 q \rho _0 (w+1)^2}.
 \end{equation}
In the above expression, we have to take into account the fact that the integration constant $\rho _0$ is, in general, a function of the parameter $q$. As we will have to use this expression to linearize the Friedmann equation around $q=0$, this dependence induces some subtleties in the derivation of our results, and we have to take into account the features of the function $\rho _0(q)$. In the following, we will leave this function completely general, neglecting terms for the order $q^2$ in the expressions we derive.

Plugging \eqref{Sol_rho_Gen} into \eqref{FriedEq}, we have, at first order in $q$,
\begin{equation}\label{LinFried}
    \frac{ \dot{S}^2}{S^2}=- \frac{k}{S^2}+
  \kappa\left[  \frac{4}{3}\frac{ \bar{\rho}_0+ q  \bar{\rho}^\dagger_0}{S^{3(1+ w)}}+  2 q (w+1)^2  \frac{   \bar{\rho}_0^2 }{ S^{6(1+w)}}\right]+\frac{\Lambda}{3}\,,
\end{equation}
where, for the sake of simplicity, we have defined 
\begin{equation}
\bar{\rho}_0=\rho_0(0) \qquad \bar{\rho}^\dagger_0=\left.\frac{d\rho_0(q)}{d q}\right|_{q=0}.
\end{equation}
Eq. \eqref{LinFried} reveals that at linear level, the MEMe model corrects the GR equation by introducing the equivalent of an effective fluid that antigravitates and has equation of state (EoS)
\begin{equation}
p=(2w+1)\rho.
\end{equation}
This result suggests the possibility of an (ultra) stiff matter era in the first stages of Cosmic history, similar to Zel'dovich's cold baryons gas model \cite{Chavanis:2014lra}. Exotic stiff EoS also appear in astrophysical settings, like neutron stars and more in general in the case of the gravitation of bosonic condensates \cite{GuerraChaves:2019foa}.   
It is clear that this term will have an important effect at early times as when $S\rightarrow0$, the term $S^{-6(1+w)}$ becomes dominant with respect to the other terms in equation \eqref{LinFried}.

\begin{figure*}[htbp]
\centering
\begin{subfigure}{.5\textwidth}
  \centering
  \includegraphics[scale=0.4]{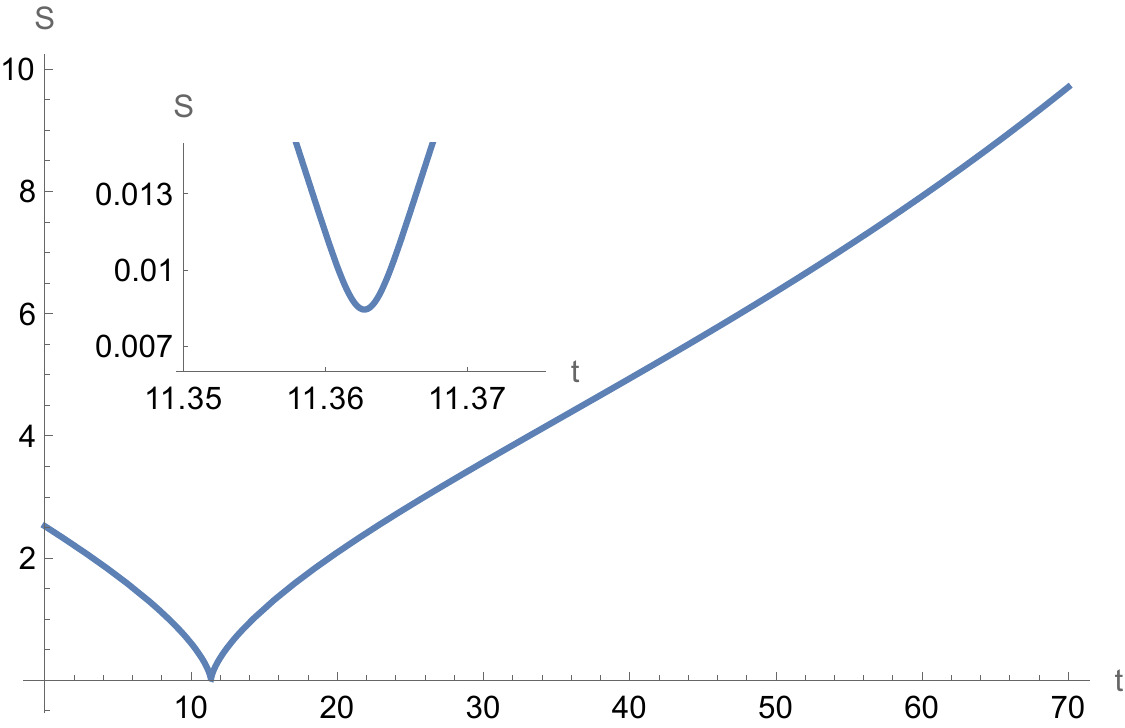}
  \caption{Plot of the scale factor \eqref{Gen_S} for $w=0$.}
\end{subfigure}%
\begin{subfigure}{.5\textwidth}
  \centering
  \includegraphics[scale=0.4]{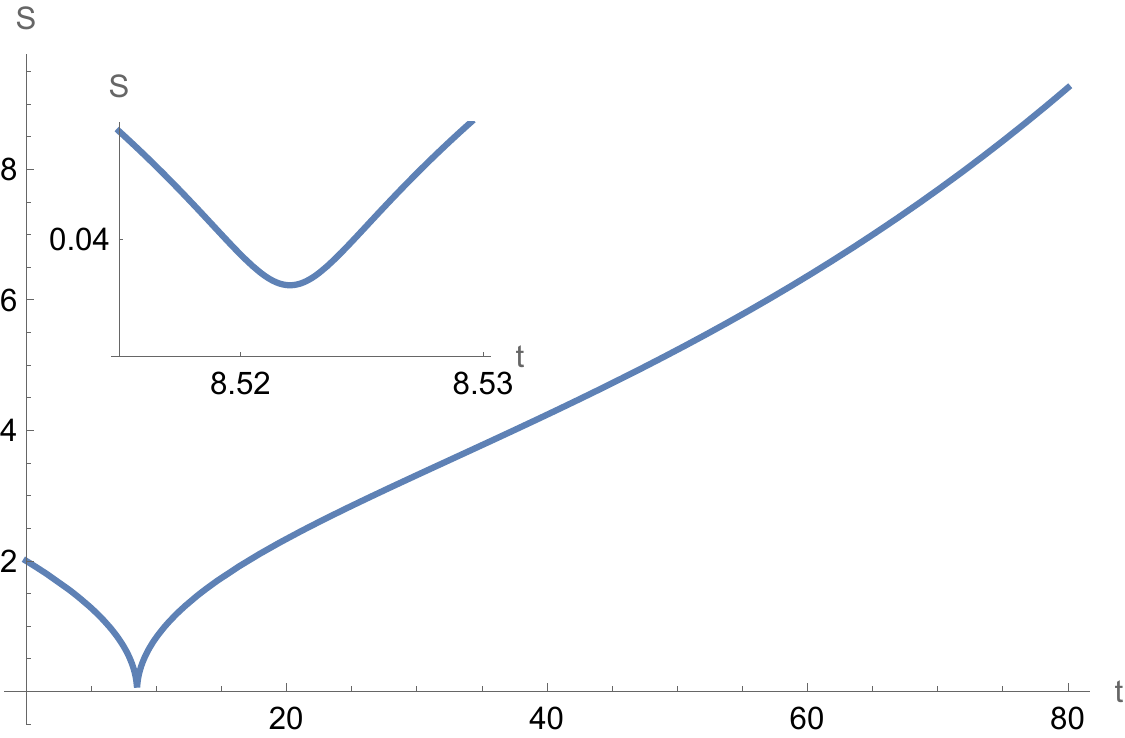} 
  \caption{Plot of the scale factor \eqref{Gen_S} for $w=1/3$.}
\end{subfigure}
\caption{Plots of the scale factor \eqref{Gen_S} for $q=-10^{-5} Gev^{-4}$, $\bar{\rho}_0=4\times10^{-2} Gev^{4} $, $\bar{\rho}^\dagger_0=2 \times10^{-5} $, and $\Lambda=10^{-3} Gev^{-2}$.}
\label{PlotS}
\end{figure*}

In the spatially flat ($k=0$) case, the general solution of Eq.\eqref{LinFried} is
\begin{equation}\label{Gen_S}
S=\left[A+B e^{-\sqrt{3\Lambda } (w+1)(t-t_0)}+C
   e^{\sqrt{3\Lambda } (w+1)(t-t_0)}\right]^{\frac{1}{3
   w+3}}
   \end{equation}
   where
   \begin{equation}
   \begin{split}\label{defABC}
   A&=-\frac{2 \kappa  \left(\bar{\rho}_0+q
  \bar{\rho}^\dagger_0\right)}{\Lambda }\\
   B&=\frac{1}{2 \Lambda  (w+1)}\\
   C&=\frac{\kappa (w+1)}{\Lambda } \left[2 \kappa  \left(\bar{\rho}_0+q
  \bar{\rho}^\dagger_0\right)^2-3 \Lambda  q \bar{\rho}_0^2(w+1)^2\right]\,.\\
   \end{split}
   \end{equation}

\begin{figure*}[htbp]
\centering
\begin{subfigure}{.5\textwidth}
  \centering
  \includegraphics[scale=0.6]{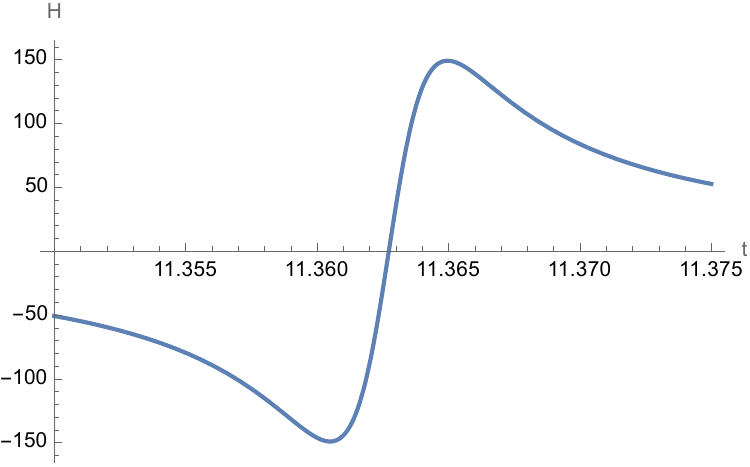}
  \caption{The Hubble parameter for the solution \eqref{Gen_S} for $w=0$}
\end{subfigure}%
\begin{subfigure}{.5\textwidth}
  \centering
  \includegraphics[scale=0.35]{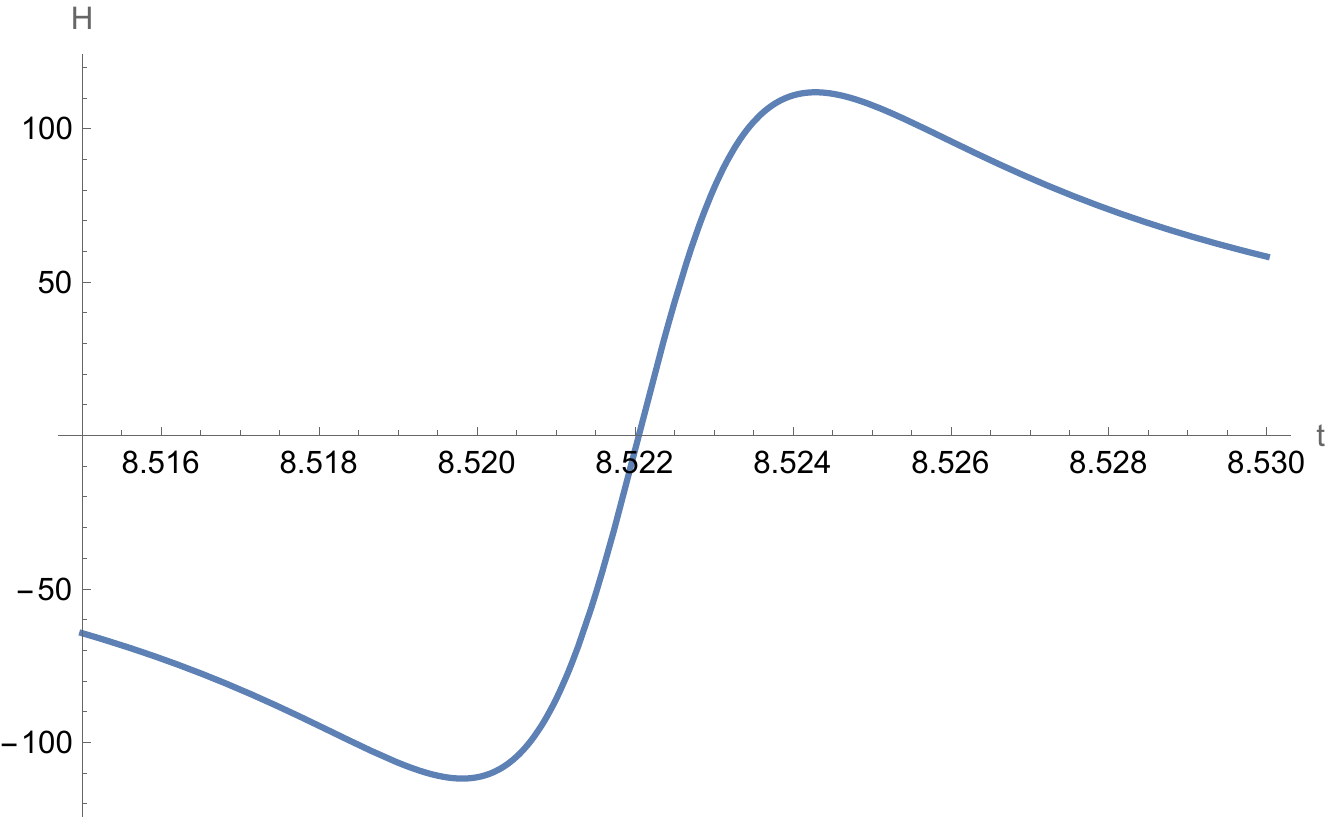}
  \caption{The Hubble parameter for the solution \eqref{Gen_S} for $w=1/3$}
\end{subfigure}
\caption{The Hubble parameter for the solution \eqref{Gen_S}. Model parameters are again set to the ones used in Figure \ref{PlotS}.}
\label{PlotH}
\end{figure*}

\begin{figure*}[htbp]
\centering
\begin{subfigure}{.5\textwidth}
  \centering
  \includegraphics[scale=0.4]{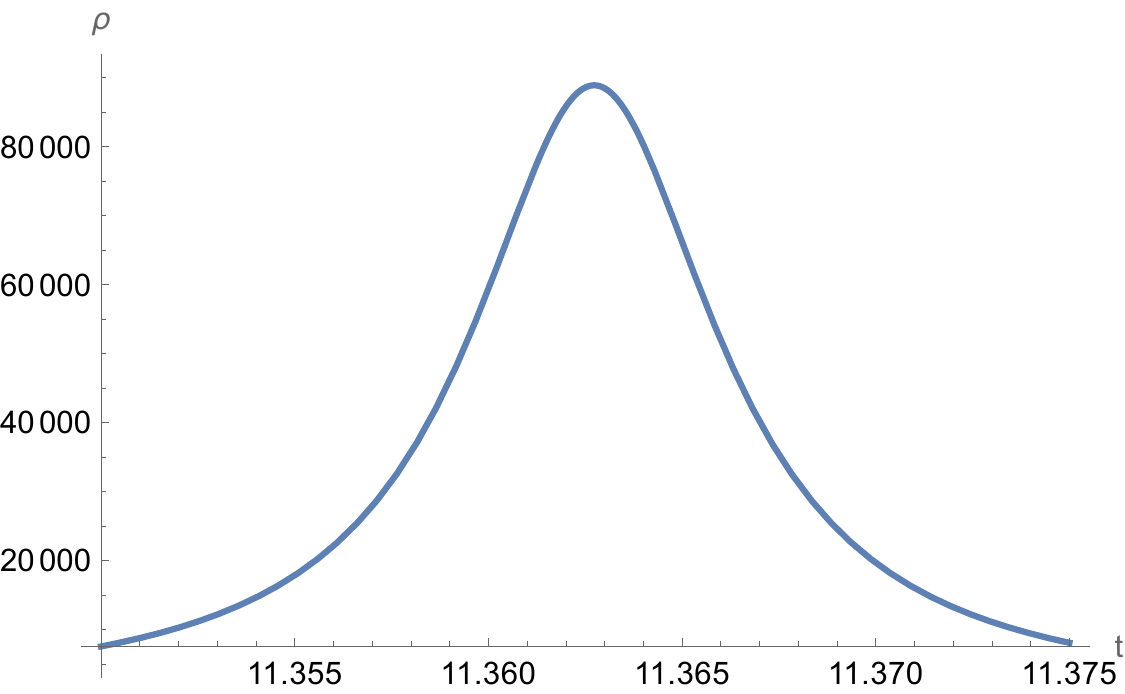}  
  \caption{Plot of the energy density  for the solution \eqref{Gen_S} for $w=0$.}
\end{subfigure}%
\begin{subfigure}{.5\textwidth}
  \centering
  \includegraphics[scale=0.4]{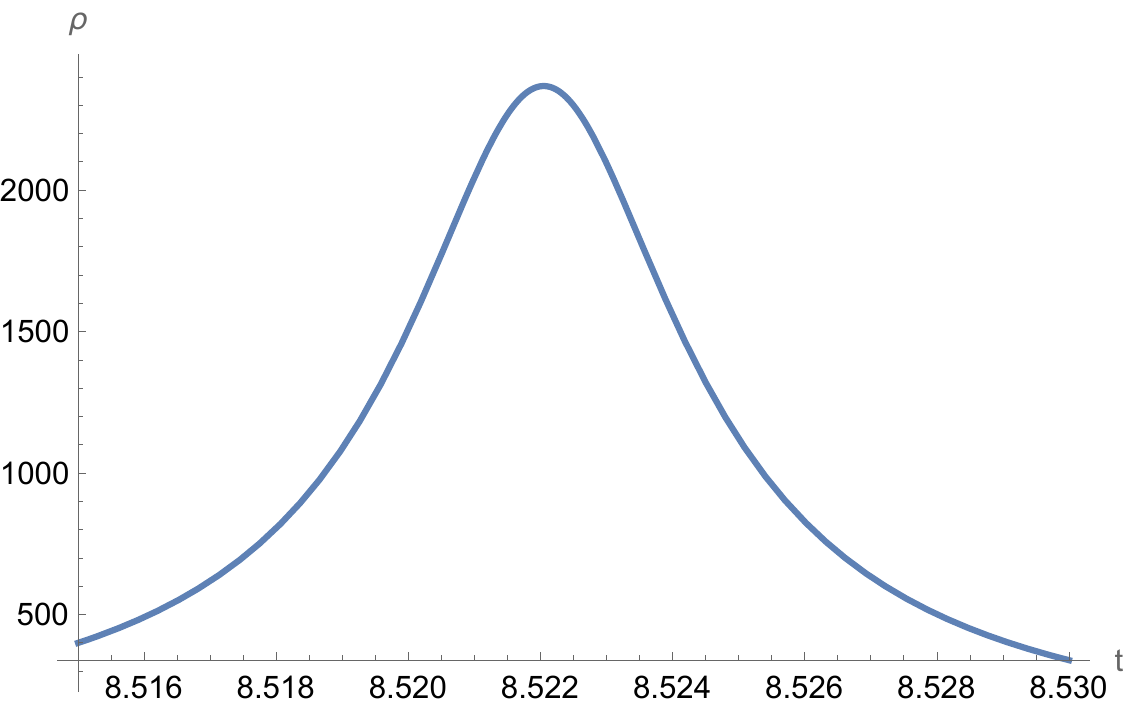}
   \caption{Plot of the energy density  for the solution \eqref{Gen_S} for $w=1/3$.}
\end{subfigure}
\caption{Plot of the energy density for the solution \eqref{Gen_S}. We adopt the same choice of model parameter as in Figure \ref{PlotS}.}
\label{PlotRho}
\end{figure*}

A plot of the behavior of the scale factor $S$ in the cases $w=0,1/3$  is given in Fig. \ref{PlotS} for $q=-10^{-5} Gev^{-4}$, $\bar{\rho}_0=4\times10^{-2} Gev^{4} $, $\bar{\rho}^\dagger_0=2 \times10^{-5} $, and $\Lambda=10^{-3} Gev^{-2}$. We adopt these parameter values, which are not compatible with observations because they are useful to visualize the behavior of $S$ and the other cosmological parameters. The solution \eqref{Gen_S}  describes a cosmic history characterized by an initial phase of contraction followed by a non-singular bounce and an expanding phase, which is at first decelerated and then accelerated. Notice that the differences between the coefficients $B$ and $C$ imply a difference in the evolution of the contracting and expanding branches. The time evolution of the Hubble parameter $H$ and the energy density $\rho$ close to the bounce are given in Figs.  \ref{PlotH} and \ref{PlotRho} using the same parameters employed for Fig. \ref{PlotS}.

It is easy to calculate the time $t_*$ at which the bounce occurs.  By setting $\dot{S}=0$, we have
\begin{equation}\label{tstar}
t_*= t_0+\frac{1}{2 \sqrt{3} \sqrt{\Lambda } (w+1)}\ln \left(\frac{C}{B}\right),
\end{equation}
which corresponds to
\begin{equation}\label{Bounce_S_Rho}
    \begin{split}
       S(t_*)= S_*&=\left(A+2 \sqrt{BC}\right)^{\frac{1}{3 (w+1)}}\,,\\
        \rho(t_*)=\rho_*&=4 \rho
   _0\left\{\left(A+2 \sqrt{BC}\right) \left[3 q \bar{\rho}_0
   (w+1)^2+4\right]\right.\\ 
  &\left.-3 q  (w+1)^2 \left(\bar{\rho}_0+q
\bar{\rho}^\dagger_0\right)\right\}^{-1}\,.
\end{split}
\end{equation}
Notice that the value of $S_*$ is not necessarily close to zero. This fact suggests that two mechanisms can contribute to the realization of the bounce in Eq. \eqref{LinFried}. One is the dominance of the term proportional to $S^{-6(1+w)}$ due to the magnitude of $S$, the other is related to the magnitude of the coefficients that multiply the $S^{-3(1+w)}$ and  $S^{-6(1+w)}$ terms. In other words, the dominance of the $S^{-6(1+w)}$ term can be achieved because $S$ is small enough or because the $S^{-3(1+w)}$ is sufficiently suppressed by its coefficient. 

Similarly, the time of onset of the dark energy era can be calculated by simply evaluating the inflection points of the scale factor:
\begin{equation}\label{tDE}
t_{DE}= t_0+\frac{1}{\sqrt{3} \sqrt{\Lambda }
   (w+1)}\ln \left( -\frac{A\pm\sqrt{A^2-4 B C}}{2 C}\right)\,,
\end{equation}
the two solutions belong to each of the two branches of the scale factor evolution. 

Hence, the results derived above suggest the following picture: during the contracting phase, the Universe experiences an accelerated regime at first, which then evolves into a decelerated collapse toward the bounce.

A remark is now in order. The bounce solution we have presented above is spatially flat. As spatial curvature becomes increasingly important in the early universe, one might consider the problem of the existence of the bounce when $k\neq0$. Unfortunately, in this case it is not immediate to get an exact solution. However, we can give a relatively easy algorithm to derive the conditions that guarantee such solutions exist.

Let us notice that a cosmic bounce can be seen as a minimum for the scale factor, and therefore, that in a bounce, $\dot{S}=0$ and $\ddot{S}>0$.
If one considers the Friedmann equation \eqref{FriedEq} and the Raychaudhuri equation \eqref{RayEq} expanded at the first order with respect to $q$, in terms of the parameters $A$, $B$ and $C$ given in \eqref{defABC}, it is possible to show that the condition $\dot{S}=0$ is equivalent to the relation
\begin{equation}\label{FriedBounce}
 S^{6 (w+1)}-2A \,S^{3 (w+1)} -k\,S^{2 (3w+2)}+A^2-4 BC =0 
\end{equation}
whereas  the condition  $ \ddot{S}>0$ requires
 \begin{equation}
\begin{split}\label{RayBounce}
&\frac{S}{6 B (w+1)}\left[2+A (3 w+1) S^{-3(w+1)}+\right.\\ &
   \left.-(3 w+2) \left(A^2-4 B C\right) S^{-6(w+1)}\right]>0.
\end{split}
\end{equation}
As an example, we can check our simple analysis in the spatially flat case ($k=0$) and a non-relativistic cosmic fluid ($w=0$). coherently with what we found in \eqref{Bounce_S_Rho}, equation \eqref{FriedBounce} has the unique solution 
\begin{equation}\label{BounceFlatCond}
 S=\sqrt[3]{A-2 \sqrt{BC}}
\end{equation}
which, once the expressions \eqref{defABC} are substituted, is always real, provided that $B C$ is positive. Introducing the definitions of $B$ and $C$, we see that this is indeed always the case. Thus, the scale factor will always have an extremum. 

In order to verify that the solution \eqref{BounceFlatCond} corresponds to a cosmic bounce, we now need to establish the sign of $\ddot{S}$. By substituting the result obtained in \eqref{BounceFlatCond} within \eqref{RayBounce}, one obtains
\begin{equation}
\frac{\ddot{S}}{S}=-\frac{\sqrt{C}}{\sqrt{B} \left(A-2\sqrt{BC}\right)^{2/3}}
\end{equation}
which is positive for all values of the parameters $q$, $\Lambda$, $\bar\rho _0$ and $\bar\rho^\dagger _0$. A similar procedure can be used to examine the case of radiation ($w=1/3$), leading to the same results. Hence, regardless of the kind of fluid that is taken into account, the term $S^{-6(1+w)}$ is always dominant in Eq. \eqref{LinFried} when $k=0$, inducing the bounce.

The procedure described above can be repeated also for the case $k\neq 0$. The resulting analysis is slightly more complicated because equation \eqref{FriedBounce} leads to a higher degree equation for $S$. As a result, one can still obtain a cosmic bounce for non-flat spatial geometries, albeit there are only very specific parameter intervals for which we recover this behavior. These conditions are too long and involved to be presented here, but the reason behind their appearance is clear. Differently from the spatially flat case, the term responsible for the bounce in \eqref{LinFried} now competes with the spatial curvature term associated with $k$. Now, the magnitude of the $S^{-2}$ term is not influenced by the parameters of the model like the $S^{-3(1+w)}$ term. Therefore, there is no guarantee that the term $S^{-6(1+w)}$ is always dominant, like in the $k=0$ case. The deciding factor is ultimately the value of the initial conditions on the energy density and its derivative with respect to $q$.

Consequently, we see that when the bounce occurs, the presence of spatial curvature does not modify the physics; in this solution, the spatial curvature term is never dominant. We, therefore, expect that the bounce will remain essentially the same as the one in the spatially flat case. For this reason, we will rely on the $k=0$ case to explore the physical properties of this scenario further.

\section{The bounce solutions in the Jordan frame}
What picture in the Jordan Frame corresponds to the one described above? In the Jordan frame, we will consider observers moving with the matter fluid and, therefore, with velocity $u_\alpha$. These observers will register a scale factor given by \cite{FengCarloni2019} 
 \begin{equation}\label{Gen_SJ}
\mathbb{S}=Y(\tau)S(\tau)= \frac{4 [1 - p(\tau)\, q]}{4 - q \,  [3 \, p(\tau) - \rho(\tau)]}S(\tau)
 \end{equation}
where $\tau$ represents the proper time for the comoving observers.  Constitutive equations allow us to write down Jordan time $\tau$ in terms of the Einstein time $t$. Indeed, from \eqref{GCA-UnitFlowField}, it is easy to obtain  that 
\begin{equation}\label{t_tau}
dt=\sqrt{-\varepsilon} d \tau=\frac{4 \, (q \,{\rho} + 1)}{4 - q \, (3 \, w - 1){\rho}} d\tau
\end{equation}
The above equation can be considered as a differential equation describing the function $\tau(t)$. By substituting  \eqref{Sol_rho_Gen} and \eqref{Gen_S} it is possible to integrate this relation exactly to obtain
\begin{equation}
\begin{split}\label{tau_t}
\tau=& t+\mathfrak{c}(t_0) +\\ & - \frac{2  q \rho_0\sqrt{ 3}}{\sqrt{\Lambda (4 BC-D) } } \tan^{-1}\left(\frac{D+2 B\, e^{\sqrt{3} \sqrt{\Lambda }(1+w)(t-t_0)}}{\sqrt{4 BC-D}}\right) \\
 D=&\left\{A+q [1-3 w (w+2)] \rho_0\right\}^2
 \end{split}
\end{equation}
where $\mathfrak{c}(t_0) $ is a function of the initial time $t_0$.

\begin{figure*}[htbp]
\centering
\begin{subfigure}{.5\textwidth}
  \centering
  \includegraphics[scale=0.4]{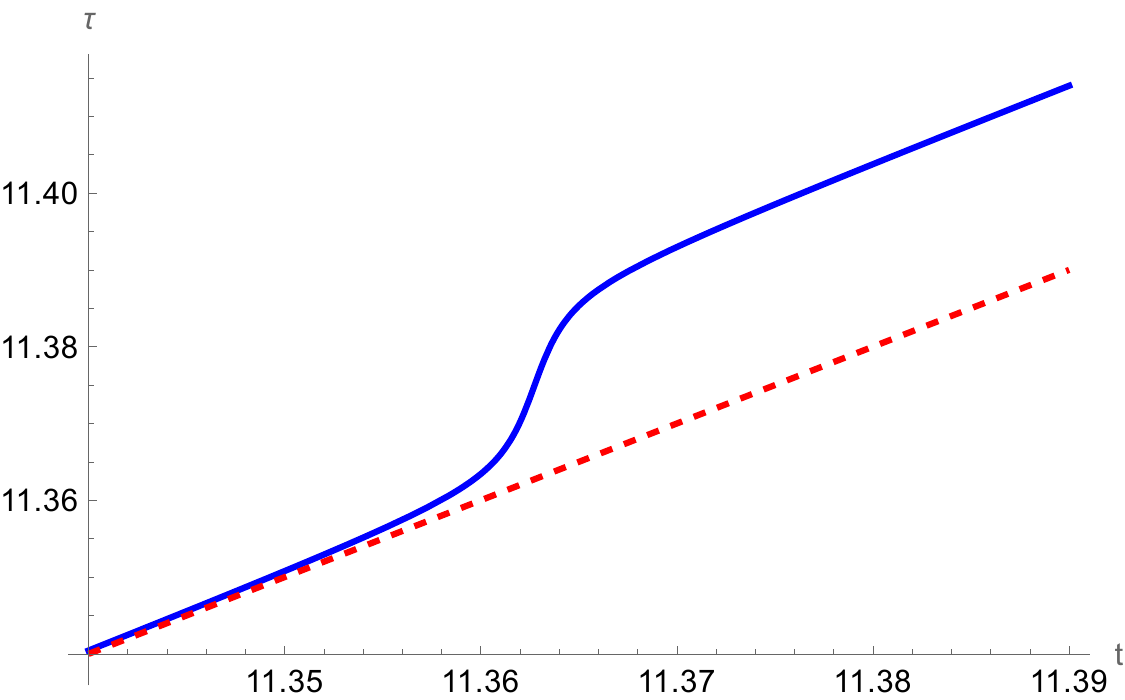} 
  \caption{Eq.(\ref{tau_t}) for $w=0$.} 
  \label{tauD_R}
\end{subfigure}%
\begin{subfigure}{.5\textwidth}
  \centering
  \includegraphics[scale=0.4]{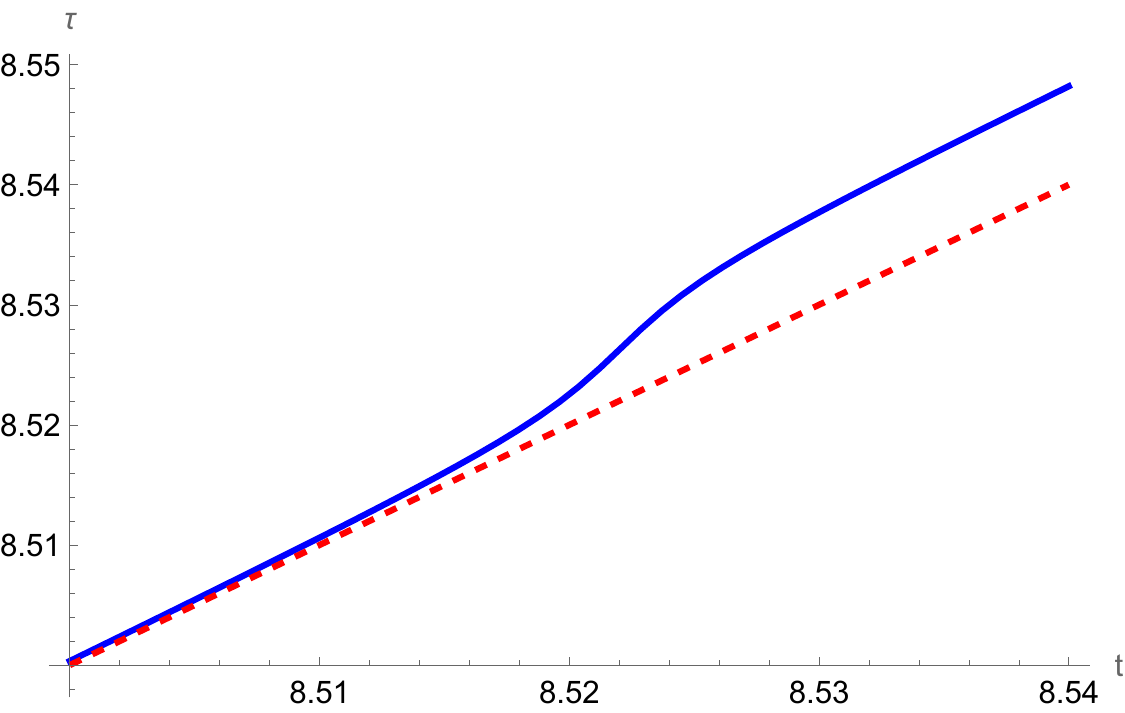}  
\caption{Eq.\eqref{tau_t} for $w=1/3$.}
 \label{t_tauD_R}
\end{subfigure}
\caption{Eq.(\ref{tau_t}) in terms of Einstein time. For reference, the dashed red line represents the synchronous case $t=\tau$. Model parameters are set as in Fig. \ref{PlotS} assuming $\rho_0=\bar{\rho}_0$.  The two time coordinates have been assumed to be zero at the same instant.}
\label{tauT}
\end{figure*}

The behaviors of \eqref{tau_t} for the dust ($w=0$) and the radiation ($w=1/3$) cases, assuming again the parameter values $q=-10^{-5} Gev^{-4}$, $\bar{\rho}_0=4\times10^{-2} Gev^{4} $, $\bar{\rho}^\dagger_0=2 \times10^{-5} $, and $\Lambda=10^{-3} Gev^{-2}$, are given in Fig.\ref{tauT} assuming, for simplicity, $\rho_0=\bar{\rho}_0$. In the following, we will always use the same parameter values.

As shown in Fig.\ref{tauD_R}, the Jordan time $\tau$ deviates from Einstein one $t$ in both cases. In particular, Jordan's time elapses faster around the bounce and thereafter remains advanced. Notice that the differences between $t$ and $\tau$ start to be evident not much earlier than the time of the bounce and then saturate to a constant very quickly.
Once the inverse time function $t=t(\tau)$ has been (numerically) calculated, it is possible to obtain the behavior of the Jordan frame cosmological variables transforming the solution (\ref{Gen_S})  using (\ref{Gen_SJ}). 

\begin{figure*}[htbp]
\centering
\begin{subfigure}{.5\textwidth}
  \centering
  \includegraphics[scale=0.4]{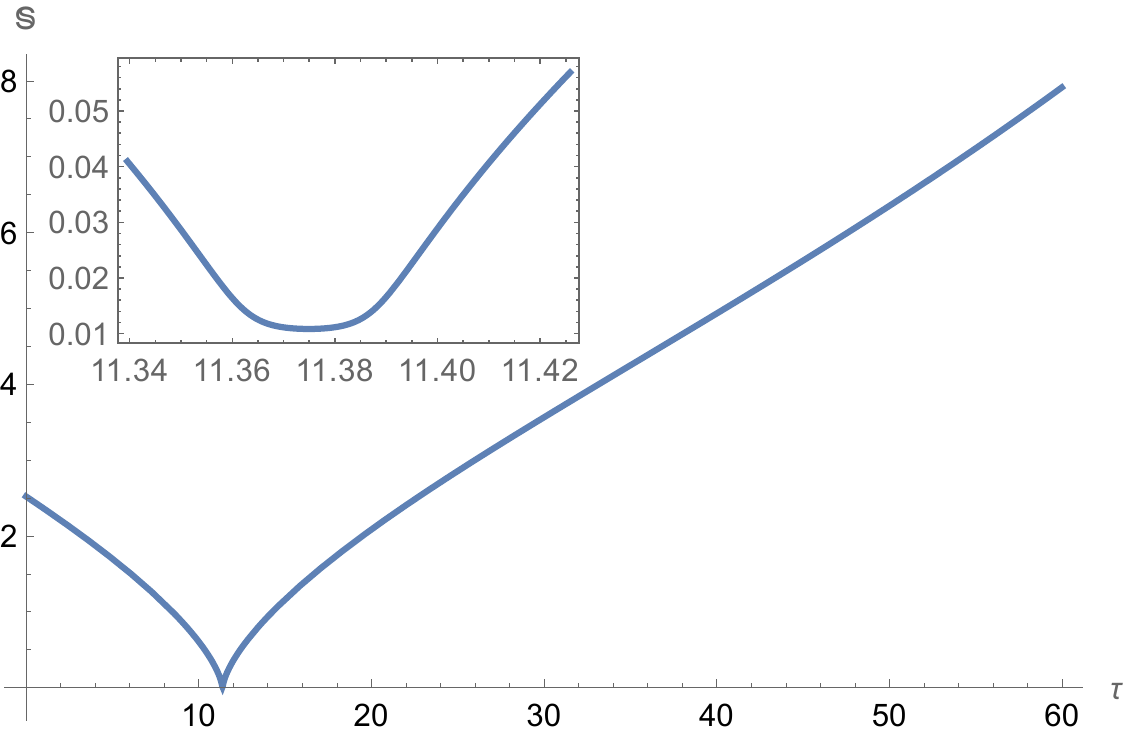} 
  \caption{The behavior of the Jordan scale factor as a function of the Jordan time for $w=0$.}
\end{subfigure}%
\begin{subfigure}{.5\textwidth}
  \centering
  \includegraphics[scale=0.4]{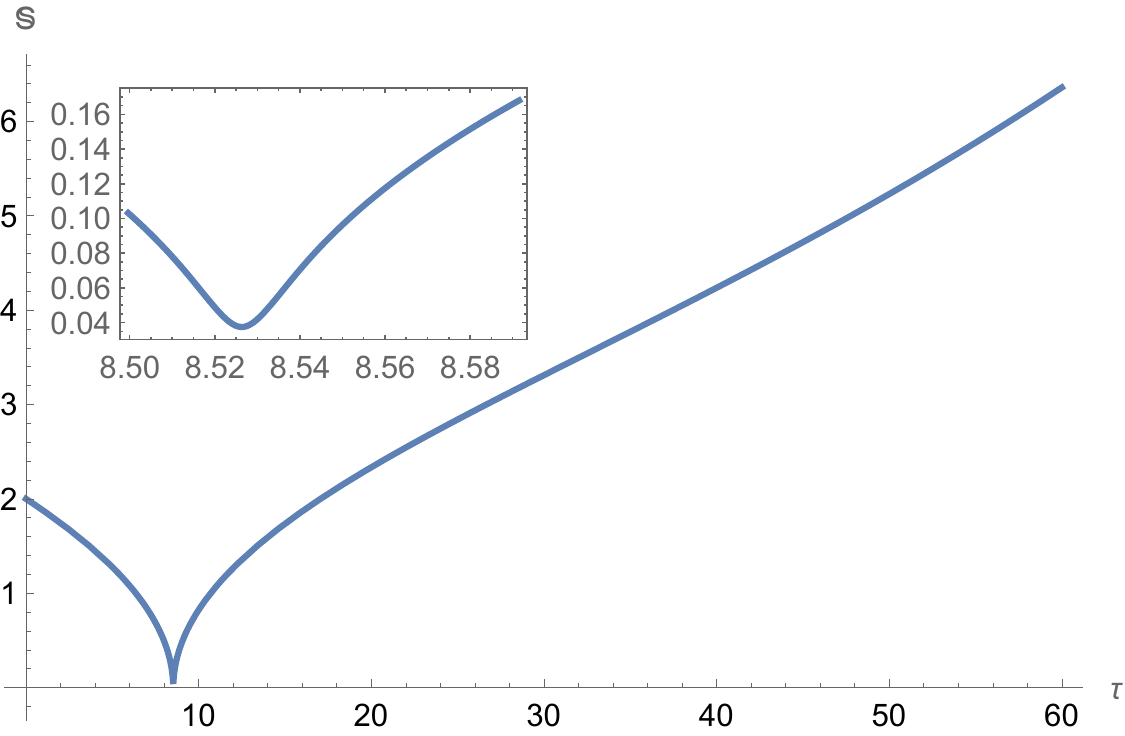}  
\caption{The behavior of the Jordan scale factor as a function of the Jordan time for $w=1/3$.}
\end{subfigure}
\caption{The behavior of the Jordan scale factor as a function of the Jordan time. }
\label{PlotScom}
\end{figure*}

\begin{figure*}[htbp]
\centering
\begin{subfigure}{.5\textwidth}
  \centering
  \includegraphics[scale=0.45]{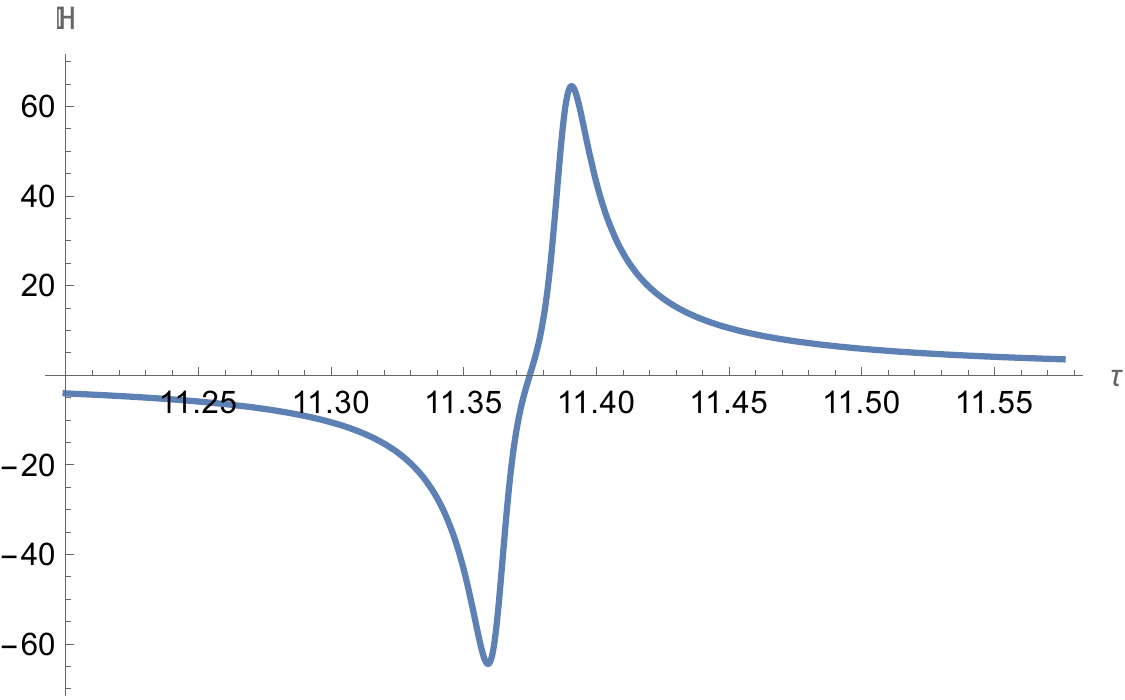}  
  \caption{Left: Plot of the Jordan Hubble parameter for dust ($w=0$).}
\end{subfigure}%
\begin{subfigure}{.5\textwidth}
  \centering
  \includegraphics[scale=0.45]{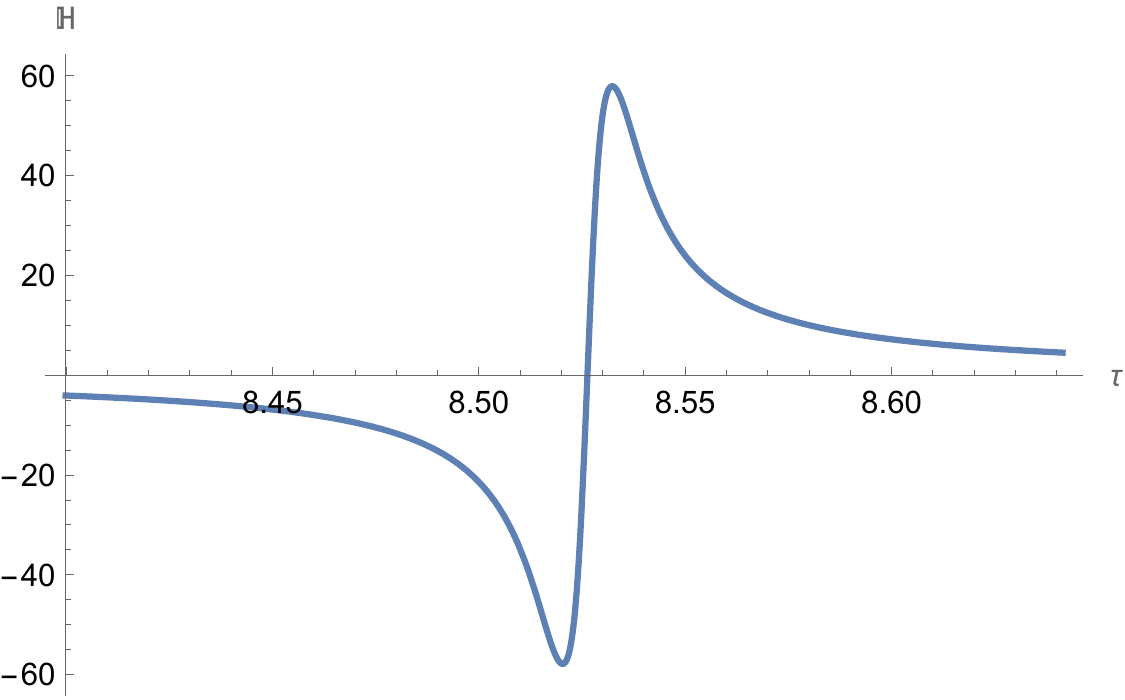} 
  \caption{Left: Plot of the Jordan Hubble parameter for dust ($w=1/3$).}
\end{subfigure}
\caption{Plot of the Jordan Hubble parameter connected with the Jordan scale factor given in (\ref{Gen_SJ}).}
\label{PlotHcom}
\end{figure*}

\begin{figure*}[htbp]
\centering
\begin{subfigure}{.5\textwidth}
  \centering
  \includegraphics[scale=0.4]{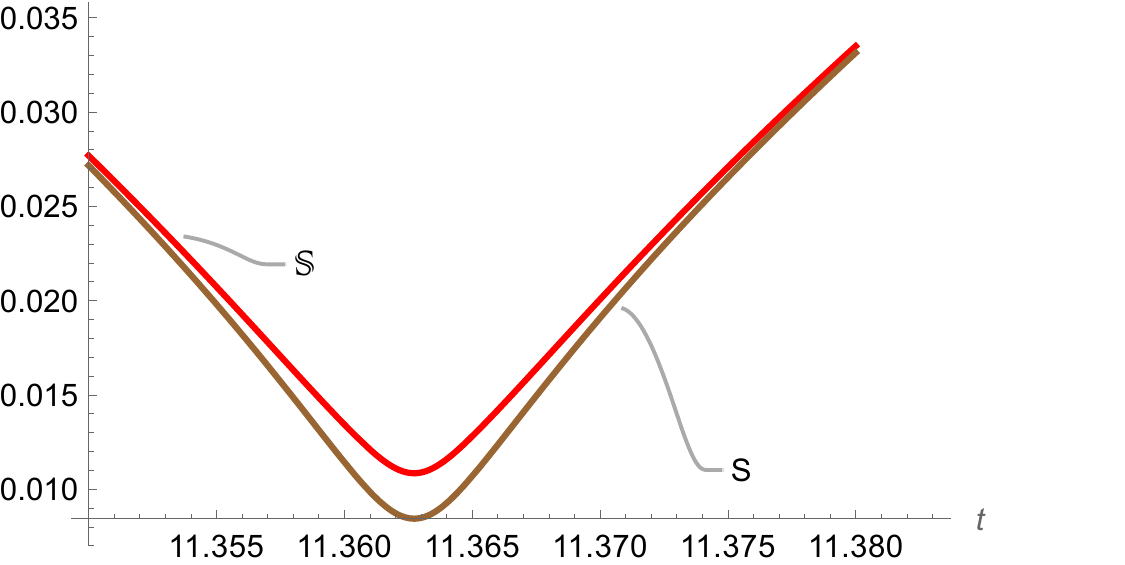}  
  \caption{A comparison of scale factor $S$ and $\mathbb{S}$ for $w=0$.}
\end{subfigure}%
\begin{subfigure}{.5\textwidth}
  \centering
  \includegraphics[scale=0.43]{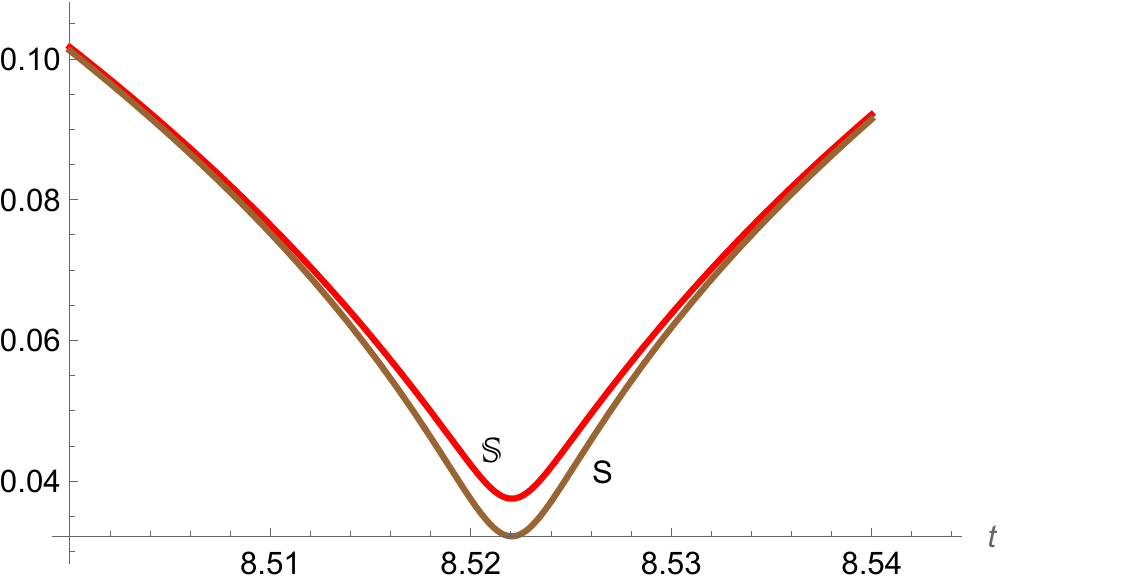}
  \caption{A comparison of scale factor $S$ and $\mathbb{S}$ for $w=1/3$.}
\end{subfigure}
\caption{A comparison of scale factor for Einstein (S) and Jordan ($\mathbb{S}$) coordinates with respect to Einstein time. }
\label{Scfactors_noncom}
\end{figure*}

\begin{figure*}[htbp]
\centering
\begin{subfigure}{.45\textwidth}
  \centering
  \includegraphics[scale=0.4]{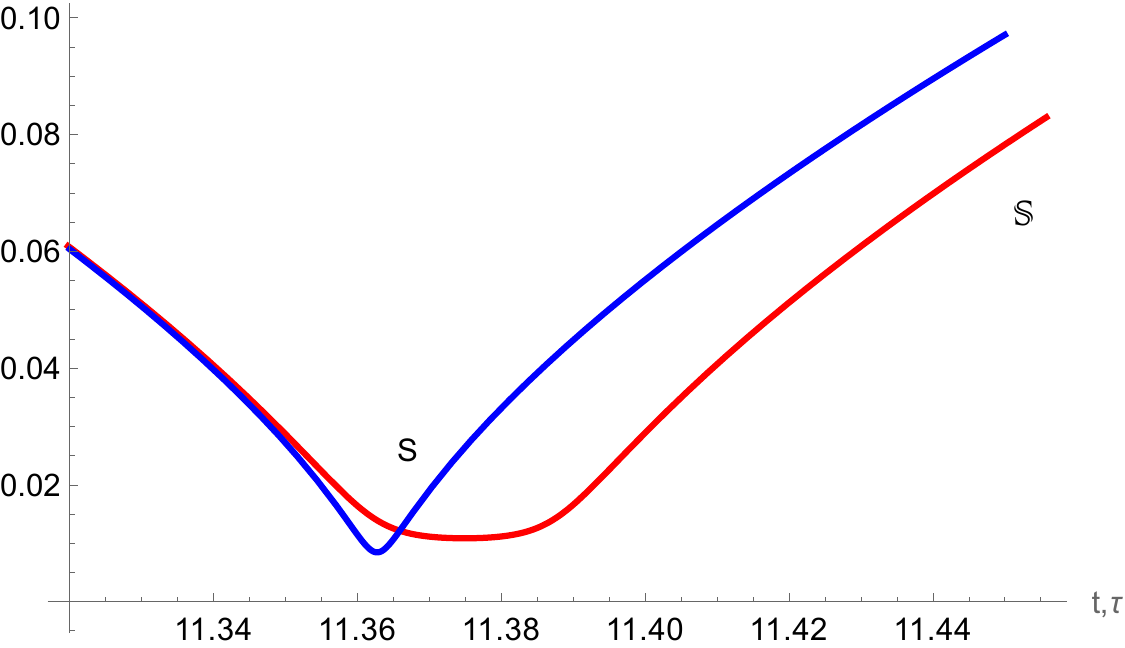}
  \caption{A comparison of scale factor $S(t)$ and $\mathbb{S}(\tau)$ for $w=0$.}
\end{subfigure}%
\begin{subfigure}{.45\textwidth}
  \centering
  \includegraphics[scale=0.4]{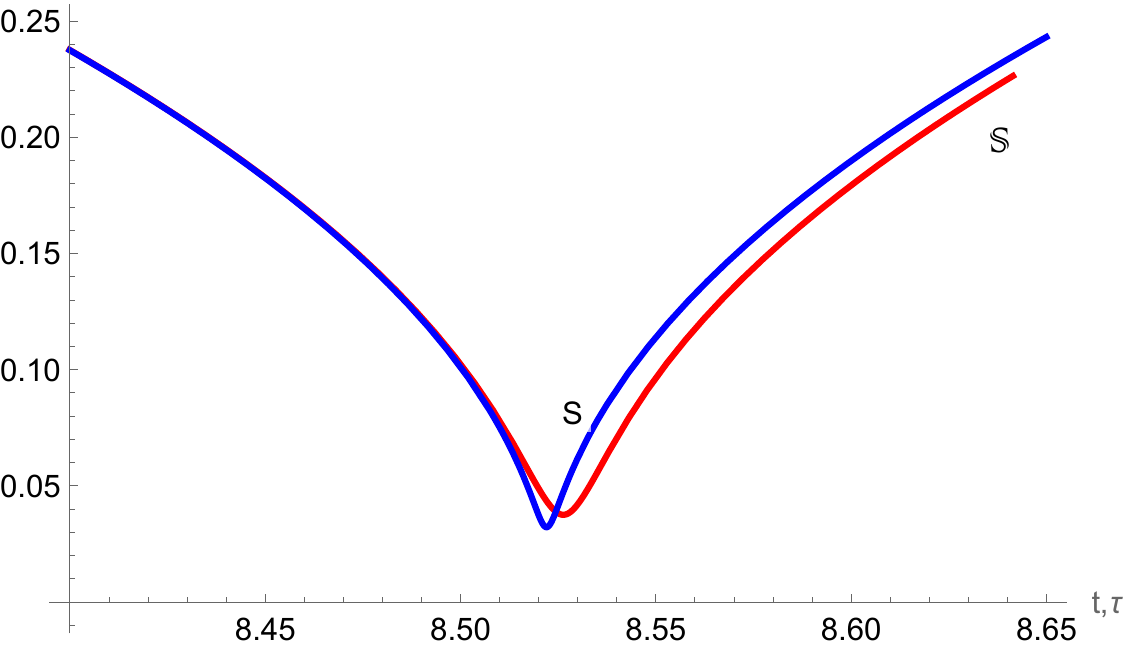} 
  \caption{A comparison of scale factor $S(t)$ and $\mathbb{S}(\tau)$ for $w=1/3$.}
\end{subfigure}
\caption{The behavior of the scale factor in the Einstein (blue line) and the Jordan (red line) frames. Both Einstein's time and Jordan's time are represented on the horizontal axis. The two time coordinates have been assumed to be zero at the same instant.  }
\label{Scfactors_ttau}
\end{figure*}

The behavior of Jordan frame scale factor $\mathbb{S}$ and Hubble parameter $\mathbb{H}$ in the case of pressureless matter and radiation are shown in Figs. \ref{PlotScom} and \ref{PlotHcom}. 
These plots show that $\mathbb{S}$ also presents a bounce, but that close to it, its evolution is different from the one of $S$. This is more evident if we factor out the difference in time coordinate by plotting $S$ and $\mathbb{S}$ in terms of the same time variable, i.e., the Einstein time $t$, see Figs.~\ref{Scfactors_noncom}  therefore expliciting the intrinsic cosmological evolution of each scale factor around the bounce. From the physical point of view, this difference is easily explained. During the collapse, as the density reaches its critical value $|q|^{-1}$, matter decouples more and more from spacetime and contracts less. As the geometry of $\mathfrak{g}, $ is constructed on the behavior of matter, and the scale factor $\mathbb{S}$ represents the variation of a volume of matter,  it is natural to expect that $\mathbb{S}>S$.
On the other hand, if we compare the scale factors in the respective proper times (by artificially synchronizing them),  we can appreciate the chronotopic difference experienced by Jordan and Einstein observers around the bounce. As one can see from Figs.\ref{Scfactors_ttau} obtained superimposing the evolution of each scale factor with respect to its proper time, the combined effects of a different cosmological evolution with a peculiar elapsing of time determines an overall distinct cosmological scenario during the collapse-expansion transition. In particular, in the Jordan frame, the bounce is shifted in time and appears shallower. 

This difference is physically significant as a shallower bounce allows matter to thermalize and smooth out inhomogeneities. In other words, in this phase, matter ``forgets'' the properties it had in the collapsing phase. The longer this loitering in the bounce, the less information will be transmitted from the contraction to the expansion phase. 

\section{Discussion and Conclusions }
 In this paper, we have presented a completely analytical bouncing cosmology based on the MEMe model. Such behavior is present in both the Einstein and the Jordan frames, albeit with slightly different characteristics. These differences, in particular the length of the bounce phase, can have a crucial effect on the imprint of the contracting phase on the expanding one. The main advantage of the model we presented is that it is wholly contained within the theory: there is no ad hoc phase that needs to be introduced to justify it, thereby limiting the predictivity of the theory. In this respect, the toy model presented, in our opinion, goes beyond the interest in the specific generalized coupling models.  Indeed, it offers a playground where many interesting processes can be analyzed quantitatively. This is the case for the evolution of cosmological perturbations throughout the bounce. In particular, the dependence of the theory on matter coupling and the presence of the different frames render very delicate the treatment of the theory at the perturbative level. For example, it might be needed to rederive the background cosmological equations as they are strictly dependent on the fluid properties.
Such analysis is well beyond the scope of this first exploratory work and will be the topic of future investigations.

The bounce scenario we have derived has the peculiarity of being asymmetric in that the contracting phase has a different profile with respect to the expanding one. This is a feature that, as far as we know, has only been found in very specific models like the VCDM \cite{Ganz:2022zgs} or cosmologies that take into full account the quantum nature of the early universe \cite{Delgado:2020htr} or in the case of Loop quantum Cosmology \cite{Chang:2018lgm}. The effect of this asymmetry and its necessity have been explored in these papers. Still, in the context of MEMe, they could lead to observable signatures that would help to test the MEMe model against cosmological observations. 

The key engine of the bounce mechanism is the combination of the decoupling of matter from ``pure'' spacetime and the behavior of the effective cosmological constant, which allows the start of a new expansion phase. This mechanism prevents, within the MEMe model, the appearance of cosmological singularities. It is natural to ask how robust this mechanism is and if, in particular, it would be possible to have a dynamical process that leads to a singularity in the Jordan frame metric while the Einstein one remains regular. The obvious realization of this scenario, given the value and sign of $q$ and $\lambda$, would be not to have a cosmological constant at all. However, because of the complicated form of the cosmological equations, it is not necessarily elementary to recognize phases in which one or the other form of matter-energy is dominant. Therefore, it is impossible to simply assume that the cosmological constant is irrelevant as customarily done in GR. Yet an investigation of this type would reveal hints of a potential connection between singularities in the Jordan and Einstein frames. 

It should also be pointed out that in this analysis, we assumed that, during the contraction, the geometry remains homogeneous and isotropic, and the matter remains a perfect fluid. These are both very strong assumptions that, in a more detailed analysis, should be softened. Although considering inhomogeneous and anisotropic metrics might complicate the behavior of matter,  the most relevant difference would undoubtedly be brought about by changes in the relation between the coupling tensor and the matter variables. These changes would dramatically modify the dynamics of the entire cosmology and might even prevent the onset of the bounce. A similar situation would arise if the matter source were composed of several fluids: the field equations should be rederived from the constitutive action, and it is not immediate to make predictions of the form of the coupling field. Naturally, a first approximation of such a generalization could be to simply consider the values of the parameter $w$ between the standard ones. For example, a mix of dust and radiation could be obtained by considering $w=1/5$, etc. An investigation by inspection, however, does not reveal any dramatic changes in the solutions we have presented above, suggesting that the bounce occurrence is not very sensible to the thermodynamics of standard matter sources. 

A final consideration concerns the consequences of the lack of dynamical formation of singularities in the context of astrophysics. Will black holes be the end result of gravitational collapse in this context? We know that in the context of the MEMe model, junction conditions might present several complications \cite{Feng:2022rga}, and these might affect the Oppenheimer-Snyder process. This fact, added to the differences in the geometrical settings, does not allow us to conclude that our results in this work imply the absence of black holes in the MEMe model. Also, in this case, dedicated studies might shed further light on the question.

\section*{Acknowledgments}
This work has been carried out in the framework of activities of the INFN Research Project QGSKY.




\end{document}